\documentclass{elsart}

\usepackage{amssymb}

\begin{document}

\begin{frontmatter}


footnotes;

\title{Second central extension in Galilean covariant field theory}


\author{C. R. Hagen}

\address{Department of Physics and Astronomy\\
University of Rochester\\
Rochester, N.Y. 14627-0171}

\begin{abstract}
The second central extension of the planar Galilei group has been
alleged to have its origin in the spin variable.  This idea is
explored here by considering local Galilean covariant field theory
for free fields of arbitrary spin.  It is shown that such systems
generally display only a trivial realization of the second central
extension.  While it is possible to realize any desired value of the
extension parameter by suitable redefinition of the boost operator,
such an approach  has no necessary connection to the spin of the
basic underlying field.
\end{abstract}


\end{frontmatter}

\section{Introduction}

It is well known that the Galilei group in two spatial dimensions has
a two dimensional central extension.  Specifically, this means that
one can expand the set of seven Galilean operators $P_i, K_i, H, J,$
and $M$ appropriate to a one dimensional central extension satisfying
the structure relations
$$[J,K_j]=i\epsilon_{ij}K_j,    [K_i,H]=0,$$
$$[J,P_i]=i\epsilon_{ij}P_j,    [K_i,K_j]=0$$
$$[J,H]=[J,M]=0,   [K_i.P_j]=i\delta_{ij}M$$
$$[P_i,P_j]=[P_i,H]=[H,M]=[K_i,M]=[P_i,M]=0$$
by the inclusion of an operator $\kappa$ which appears only in the
modified commutator
$$[K_i,K_j]=i\epsilon_{ij}\kappa$$
and commutes with all other operators of the theory.  The operators
$M$ and $\kappa$ thus provide a two dimensional central
extension.

While the operator $M$ has a well understood role in Galilean
invariant theories, the physical origin of $\kappa$ is less clear.
Duval and Horv\'{a}thy \cite{Duval} have recently considered this
issue as have
Jackiw and Nair \cite{Jackiw}.  Although the latter claim to have
established its identification with the spin of the relevant
particle, there are a
number of issues which need to be raised concerning their derivation.
In particular their Eq.(5) has the form
$$J_a=\epsilon_{abc}x^bp^c+{sc^2p_a\over (p^2)^{1\over 2}},$$
an equation which implies that the spin has the dimension of $1/c^2$
rather than the dimensionless form expected in units in which
$\hbar=1$.  This is also the basis for their claim that in the limit
of $c\to \infty$ the second term on the rhs of the above equation
dominates.  Physically, this is an odd result since it seems to
assert that the spin angular momentum necessarily dominates over
orbital angular momentum in the large $c$ limit.  Both of these
issues clearly have their origin in the occurrence of seemingly
anomalous factors of $c$ in the derivation of \cite{Jackiw}.

One of the most forthright approaches to an objective assessment of
the validity of the Jackiw-Nair claim is to examine this issue in the
context of specific models. To this end one seeks (preferably simple)
examples in which the spin
degree of freedom does indeed give rise to the second central
extension parameter $\kappa$.  This is certainly feasible within the
context of local Galilean covariant field theory describing particles
of nonzero spin.  Clearly, if one can construct such theories
corresponding to a $\kappa$ which either vanishes or is unrelated to
spin, the question will be unambiguously settled.  In the following
section the Galilean field theory of a free spin one-half particle is
constructed and shown to have a vanishing second central extension
coefficient.  In {\bf 3} this is extended to the case of arbitrary
spin, with similar results being obtained.

\section{Galilean Spin One-Half}

The Galilean covariant wave equation for a spin one-half particle was
constructed by L\'{e}vy-Leblond [3].  His result in three spatial
dimensions requires a four-component spinor just as in the case of
the corresponding Dirac equation.  Another feature which it shares
with the Dirac equation is the prediction of $g=2$ for the $g$-factor.
In the case of two spatial dimensions the four-component spinor
equation separates into two two-component equations.  Denoting the
two spin components $\pm{1\over 2}$ by the spin parameter $s$ with
$s=\pm 1$ one can write the Lagrangian for a spin one-half particle as
$$\mathcal{L}=\psi^\dagger \left[{1\over 2}
(1+\sigma_3)i{\partial\over \partial t}
-i{\bf \sigma}\cdot{\bf \nabla} +m(1-\sigma_3)\right]\psi$$
where ${\bf \sigma}=(\sigma_1,s\sigma_2)$ and $\sigma_i, i=1,2,3$
denote the usual Pauli matrices.  Upon making the identification
$\psi_1=\phi$, $\psi_2=\chi$, one obtains the equation of motion for
the independent component $\phi$
\begin{equation}
E\phi+p_{\mp}\chi=0
\end{equation}
with the dependent component $\chi$ given by

\begin{equation}
       2m\chi+p_{\pm}\phi=0
\end{equation}
where $E=i{\partial\over \partial t}, p_{\pm}=-i(\nabla_1\pm is\nabla_2)$.

The invariance of the Lagrangian under rotations by an angle
$\varphi$ readily follows from the transformation law
\begin{equation}
       \psi'({\bf x'}) = \exp [i{s\over 2}\sigma_3\varphi] \psi({\bf x})
\end{equation}
appropriate to a particle of spin $s/2$.  The corresponding result
for Galilean boosts with boost parameters $v_i$ is readily verified
using the spinor transformation prescription
\begin{equation}
\psi'({\bf x'},t') =
[1-{1\over4}(\sigma_1-i\sigma_2)v_{\pm}]\exp[im{\bf v}\cdot{\bf x}+
(i/2)mv^2t]\psi({\bf x},t) .
\end{equation}
Corresponding to this statement of Galilean covariance is the
existence of the local conservation law
\begin{eqnarray*}
\nabla_j & \; & \left[\psi^\dagger
[-mx_i\sigma_j-it{1\over 2}\sigma_j\nabla_i-{s\over 4}
\epsilon_{ij}(1+\sigma_3)]\psi+i{1\over
2}\nabla_i\psi^\dagger\sigma_j\psi \right] \\
&  & + {\partial\over \partial t} \left[ \psi^\dagger{1\over 2} (1+\sigma_3)
(mx_i+it{1\over 2}\nabla_i)\psi-it{1\over
2}\nabla_i\psi^\dagger{1\over 2}(1+\sigma_3)\psi\right]  = 0
\end{eqnarray*}
and the generator of Galilean boosts
\begin{equation}
K_i=m\int x_id^2x \phi^\dagger\phi-tP_i
\end{equation}
where $P_i$ is the momentum operator
\begin{equation}
P_i=-i\int d^2x\phi^\dagger\nabla_i\phi.
\end{equation}
     From this explicit form for the Galilean boost operators and the
equal time anticommutation relation
\begin{equation}
\{\phi({\bf x}, \phi^\dagger({\bf x'})\}=\delta({\bf x}-{\bf x'})
\end{equation}
it is readily inferred that $[K_i,K_j]=0$.  This simple model thus
demonstrates that it is indeed possible to construct a nonzero spin
field theory which yields a trivial result for the second extension
parameter.

\section{Extension to Arbitrary Spin}

There is a standard tool available by which one can extend the
results of the preceding section to the case of arbitrary integral or
half-integral spin.  This is the multispinor approach in which one
describes a particle of spin $Ns/2$ by means of a totally symmetric
multispinor of rank $N$.  This approach has been successfully
implemented \cite{Hagen} in the calculation of the $g$-factor of a
Galilean particle in three spatial dimensions, and is applied even
more easily to the case at hand.  To this end one introduces the
multispinor $\psi_{a_1a_2 \ldots a_N}$
which transforms according to Eqs.(3) and (4) in each index.  Upon
noting that the matrix $\Gamma\equiv{1\over 2}(1+\sigma_3)$ is
invariant under such transformations one infers that the appropriate
Lagrangian is
\begin{equation}
\mathcal{L}={1\over N}\psi_{a_1...a_N}
\sum^N_{i=1}\Gamma_{a_1a_1'}...\Gamma_{a_{i-1}a_{i-1}'}
G_{a_ia_i'}\Gamma_{a_{i+1}a_{i+1}'}...\Gamma_{a_Na_N'}\psi_{a_1'...a_N'}
\end{equation}
where
$$G\equiv i\Gamma E+\sigma\cdot p +m(1-\sigma_3).$$

Particularly in the case of two spatial dimensions there is a great
deal of redundancy in the equations implied by the Lagrangian (8).
Specifically, because of the fact that $\Gamma$ is only nonvanishing
between upper components (i.e., $a_i=1$), there are only two distinct
equations implied by (8).  These equations are easily seen to involve
only the components $\psi_ {11...1}$ and $\psi_{21...1}$.  By
renaming them $\phi$ and $\chi$ respectively, one readily obtains
Eqs. (1) and (2) as well as the generalization of Eq. (3)
\begin{equation}
\psi^{'}({\bf x'}) = \exp [i(N-1+\sigma_3)s/2] \psi({\bf x})
\end{equation}
which are thus seen to describe Galilean particles of
spin $Ns/2$ \cite{gfactor}.  It immediately follows from this
notational change that
the expressions for the Galilean boost operators $K_i$ for spin
$Ns/2$ are identical to those given by  Eqs. (5) and (6).  This
together with the anticommutator (7) suffices to establish that the
boost operators commute with each other, thereby demonstrating that
only trivial realizations of the second central extension are
realized in this model.

While the multispinor approach allows only integral $N$ (i.e.,
integral and half-integral values of the spin), it is not difficult
to prescribe an extension of the two component spinor model for spin
$Ns/2$ to arbitrary spin merely by changing the definition of the
angular momentum operator.  In particular this can be achieved by the
replacement
$$J\to J+\lambda M/m$$
where $\lambda$ is any real number and $M$ is the mass operator
$$M=m\int d^2x \phi^\dagger\phi.$$
This allows one to assign any value to the spin while leaving intact
the structure relations of the Galilei group.  The second central
extension parameter is again zero for this extended model.

\section{Conclusion}

In this work it has been shown that in a very wide class of arbitrary
spin  Galilean field theories there is no connection between the spin
and the second central extension of the Galilei group.  While the
extension to arbitrary spin is significant, it is perhaps true that
the most relevant result is that for spin one-half.  This is because
in 2+1 dimensions the Dirac equation goes over unambiguously to the
L\'{e}vy-Leblond spin one-half equation leaving no possibility for a
misidentification of the spin variable as one passes from the special
relativistic case to the Galilean relativistic one.

On the other hand the fact that spin does not seem capable of
yielding a nonzero $\kappa$ certainly does not mean that such
theories cannot be constructed.  It has in fact been noted
\cite{Hagen2} that one can always proceed from a nonzero $\kappa$
theory to one with $\kappa=0$ merely by defining a new Galilean boost
operator $K_i'$ by the prescription
\begin{equation}
K_i'\equiv K_i+{1\over 2}\kappa M^{-1}\epsilon_{ij}P_j.
\end{equation}
Such a redefinition takes one to a new set of Galilean operators with
a trivial second central extension.  Naturally this operation can be
carried out in either direction, and one thus easily obtains a
$\kappa\neq 0$ theory from the class of $\kappa=0$ models considered
in this work merely by implementing the inverse of Eq. (10).  The
second central exstension of the two-dimensional Galilei is thus more
closely related to the possibility of including translations in the
boost operators than it is to the particle spin.

\label{}



\section*{Acknowledgement}
Correspondence on this subject with P. A. Horv\'{a}thy is gratefully
acknowledged.  This work is supported in part by the U.S. Department
of Energy Grant
No.DE-FG02-91ER40685.


\begin{thebibliography}{99}
\bibitem{Duval} C. Duval and P. A. Horv\'{a}thy, Phys.Lett. B 479 (2000) 284.
\bibitem{Jackiw} R. Jackiw and V. P. Nair, Phys.Lett. B 480 (2000) 237.
\bibitem{Leblond} J. M. L\'{e}vy-Leblond, Commun. Math. Phys. 6 (1967) 286.
\bibitem{Hagen} C. R. Hagen and W. J. Hurley, Phys. Rev. Lett. 24 (1970) 1381.
\bibitem{gfactor} It is readily established following the approach of
\cite{Hagen} that the $g$-factor in this theory is $1/S$ ($S\equiv
N/2$) just as it is in the case of minimal Galilean theories in three
spatial dimensions.
\bibitem{Hagen2} C. R. Hagen, Phys. Rev. 31 (1985) 848.





\end{thebibliography}
\end{document}